\documentclass[%
 aip,
 amsmath,amssymb,
 reprint,%
]{revtex4-1}

\usepackage{graphicx}% Include figure files
\usepackage{dcolumn}% Align table columns on decimal point
\usepackage{bm}% bold math
\usepackage[utf8]{inputenc}
\usepackage[T1]{fontenc}
\usepackage{mathptmx}
\usepackage{etoolbox}

\makeatletter
\def\@email#1#2{%
 \endgroup
 \patchcmd{\titleblock@produce}
  {\frontmatter@RRAPformat}
  {\frontmatter@RRAPformat{\produce@RRAP{*#1\href{mailto:#2}{#2}}}\frontmatter@RRAPformat}
  {}{}
}%
\makeatother
\begin{document}

\preprint{AIP/123-QED}

\title[Differential pumping for kHz LWFA]{Differential pumping for kHz operation of a Laser Wakefield accelerator based on a continuously flowing Hydrogen gas jet}

\author{J. Monzac}
\email{josephine.monzac@ensta-paris.fr}
\affiliation{ LOA, CNRS, \'Ecole Polytechnique, ENSTA Paris, Institut Polytechnique de Paris, Palaiseau, France}
\author{S. Smartsev}
\affiliation{ LOA, CNRS, \'Ecole Polytechnique, ENSTA Paris, Institut Polytechnique de Paris, Palaiseau, France}
\author{J. Huijts}%
 \affiliation{ LOA, CNRS, \'Ecole Polytechnique, ENSTA Paris, Institut Polytechnique de Paris, Palaiseau, France}
 \author{L. Rovige}
\affiliation{ LOA, CNRS, \'Ecole Polytechnique, ENSTA Paris, Institut Polytechnique de Paris, Palaiseau, France}

 \author{I.A. Andriyash}
\affiliation{ LOA, CNRS, \'Ecole Polytechnique, ENSTA Paris, Institut Polytechnique de Paris, Palaiseau, France}
 \author{A. Vernier}
\affiliation{ LOA, CNRS, \'Ecole Polytechnique, ENSTA Paris, Institut Polytechnique de Paris, Palaiseau, France}
\author{V. Tomkus}
\affiliation{Center for Physical Sciences and Technology, Savanoriu Ave. 231, LT-02300, Vilnius, Lithuania}
\author{V. Girdauskas}
\affiliation{Center for Physical Sciences and Technology, Savanoriu Ave. 231, LT-02300, Vilnius, Lithuania}
\affiliation{Vytautas Magnus University, K. Donelaicio St. 58. LT-44248, Kaunas, Lithuania}
 \author{G. Raciukaitis}
 \affiliation{Center for Physical Sciences and Technology, Savanoriu Ave. 231, LT-02300, Vilnius, Lithuania}
 \author{M. Mackevi\v ci\=ut\.{e}}
\affiliation{Center for Physical Sciences and Technology, Savanoriu Ave. 231, LT-02300, Vilnius, Lithuania}
\author{V. Stankevic}
\affiliation{Center for Physical Sciences and Technology, Savanoriu Ave. 231, LT-02300, Vilnius, Lithuania}
  \author{A. Cavagna}
 \affiliation{ LOA, CNRS, \'Ecole Polytechnique, ENSTA Paris, Institut Polytechnique de Paris, Palaiseau, France}
 \author{J. Kaur}
 \affiliation{ LOA, CNRS, \'Ecole Polytechnique, ENSTA Paris, Institut Polytechnique de Paris, Palaiseau, France}
 \author{A. Kalouguine}
 \affiliation{ LOA, CNRS, \'Ecole Polytechnique, ENSTA Paris, Institut Polytechnique de Paris, Palaiseau, France}
\author{R. Lopez-Martens}
\affiliation{ LOA, CNRS, \'Ecole Polytechnique, ENSTA Paris, Institut Polytechnique de Paris, Palaiseau, France}
\author{J. Faure}
\affiliation{ LOA, CNRS, \'Ecole Polytechnique, ENSTA Paris, Institut Polytechnique de Paris, Palaiseau, France}

\date{\today}% It is always \today, today,
             %  but any date may be explicitly specified

\begin{abstract}
Laser-Wakefield Accelerators (LWFA) running at kHz repetition rates hold great potential for applications. They typically operate with low-energy, highly compressed laser pulses focused in high-pressure gas targets. Experiments have shown that the best-quality electron beams are achieved using Hydrogen gas targets. However, continuous operation with Hydrogen requires a dedicated pumping system. In this work, we present a method for designing a differential pumping system, which we successfully implemented in our experiments. This enabled the first demonstration of continuous operation of a kHz LWFA using a high-pressure Hydrogen gas jet. The system effectively maintained a pressure below $3 \times 10^{-4}\,$mbar, even with a free-flowing gas jet operating at 140 bar backing pressure. Numerical fluid dynamics and optical simulations were used to guide and validate the system’s design.
\end{abstract}

\maketitle

\section{\label{sec:level1}Introduction}

Laser wakefield acceleration (LWFA) is a process that enables the generation and acceleration of electron beams to relativistic energies over very short distances \cite{tajima_laser_1979}. An ultrashort laser pulse is focused into a plasma, and drives a large density amplitude plasma wave in which plasma electrons can be trapped and accelerated. Since the plasma is already ionized, the electric field of the plasma wave is not subject to the breakdown limit that exists in conventional radiofrequency (RF) cavities. Thus the accelerating field can reach amplitudes up to 4 orders of magnitude higher than in conventional RF cavities: acceleration occurs over much shorter distances, paving the way for compact accelerators.\\
In LWFA, the laser first propagates through a vacuum before being focused into a plasma, where the acceleration process occurs. The plasma is generated as the laser ionizes the gas supplied by the target system. Various types of gas targets are used in LWFA. Each type is designed to balance the required plasma density while minimizing gas leakage, which increases the pressure in the vacuum chamber where the laser propagates. Here are the most commonly used target systems: (i) gas jets: a supersonic gas flow expands from a nozzle into the vacuum chamber \cite{semushin_high_2001, schmid_supersonic_2012}, (ii) gas cells: the gas is enclosed in a cell, creating an interaction volume in which the laser propagates through two apertures at the cell’s entrance and exit \cite{osterhoff_generation_2008}. These apertures must be small enough to minimize gas leakage in the main chamber, but large enough to avoid material ablation caused by the laser which would in turn negatively affect the accelerator performance. In order to mitigate the leaks, the gas cells operate at lower gas pressures, resulting in lower plasma densities compared to gas jets. Additionally, some gas cells utilize differential pumping, where the gas is removed from the cell simultaneously as it is introduced \cite{kirchen_optimal_2021, drobniak_two-chamber_2023}. A last commonly used target system is the capillary discharge target, where the plasma is generated by a high-voltage discharge within a capillary filled with gas. There, the laser propagates through the guided mode of the plasma channel formed during plasma expansion and interaction with the capillary walls \cite{leemans_gev_2006}. This approach is relevant when long acceleration distances are needed (in the range of $1\,$cm), but the entrance capillary aperture can be damaged by high-intensity lasers. It is worth noticing that similar target systems are also being employed in the field of nuclear physics. Notably, some of their experiments make use of gas jets combined with differential pumping techniques, though these systems generally operate at much lower densities than those in the present study \cite{ulbricht_high_1972, bittner_windowless_1979, treichel_differentially_1983, shapira_hhirf_1985}. Early-stage designs and implementations of such differential pumping systems have been explored in the field of high harmonic generation \cite{sayrac_pressure_2015} or LWFA \cite{bohlen_stability_2022}.\\
Advances in laser technology, along with the development of the target systems, have led to tremendous progress in the laser-plasma accelerator performances. Today, LWFA driven by 100 TW to PW scale laser systems can produce electron beams with energies ranging from $100\,$MeV to a few GeV over a few centimeters \cite{mangles_monoenergetic_2004, geddes_high-quality_2004, faure_laserplasma_2004, kim_stable_2017, oubrerie_controlled_2022, miao_multi-gev_2022}. To date, the highest energy ever reached is 8 GeV using plasma discharges technology \cite{gonsalves_petawatt_2019}.\\
While pushing the energy frontier is important for high-energy physics, another line of research focuses on the development of high repetition-rate (100 Hz - 1 kHz) LWFAs, which holds significant potential for enhancing stability and enabling the acquisition of large datasets, essential for robust statistical analysis. Until recently, kHz laser systems were restricted to a few millijoules per pulse. To fulfill the conditions for laser-wakefield acceleration using such energies, it is necessary to tightly focus and strongly compress the laser pulse, and to work with very high electron densities in the plasma on the order of $n_e \sim 10^{20}\,\textrm{cm}^{-3}\ $ \cite{faure_review_2018, guenot_relativistic_2017, gustas_high-charge_2018, salehi_mev_2017}. The high electron density imposes the use of free-flowing gas jets, which can provide such densities while refreshing the target quickly enough to handle kHz operation. The continuous operation of LWFA at 1 kHz has been demonstrated using Nitrogen \cite{guenot_relativistic_2017, rovige_demonstration_2020}. However, recent experimental work shows that  using light gases such as Hydrogen is beneficial to LWFA and that higher electron energy and better spatial beam properties can be reached \cite{salehi_laser-accelerated_2021}.\\
Regarding vacuum conditions, utilizing a nitrogen plasma is convenient: each $\mathrm{N_2}$ molecule provides 10 electrons, enabling a backing pressure of only 20 to 30 bar in the gas jets to achieve the required plasma density. Additionally, heavy gases like nitrogen are easier to pump than light gases \cite{gaede_diffusion_1915, becker_turbomolecular_1966}. However, ionization effects can significantly distort the laser pulse in $\mathrm{N_2}$ as different levels are ionizes at different positions within the laser pulse, causing a decrease in accelerator performance. To counter this, utilizing a light gas such as hydrogen is necessary \cite{monzac_optical_2024}. So far, kHz-scale experiments using an $\mathrm{H_2}$  plasma have been limited to burst mode operation as pumping $\mathrm{H_2}$ poses significant challenges \cite{salehi_mev_2017, salehi_laser-accelerated_2021}. When a free-flowing gas jet of $\mathrm{H_2}$  is used at high pressure, it results in considerable gas loading in the vacuum chamber where the experiment is conducted. This phenomenon was observed by Salehi et al. \cite{salehi_laser-accelerated_2021}, who reported a rapid pressure increase in the main chamber, rising from $2.6 \times 10^{-2}\,$mbar to $2.0 \times 10^{-1}\,$mbar in just 1 second when using a free-flowing gas jet with a $\sim 150\,\text{\textmu}$m FWHM aperture and a backing pressure of $\sim 30\,$bar. They operated their accelerator in burst mode to avoid gas loading in their experiments.\\
This paper describes the design, implementation, and testing of a differential pumping scheme for a continuously flowing gas nozzle used with light gases in a kHz laser-plasma accelerator. It is structured as follows: Section II introduces the physical principles underpinning the design of the differential pumping system, as well as a thorough analysis demonstrating its necessity and effectiveness. In Section III, we conduct fluid simulations and laser propagation simulations to verify the absence of laser distortion caused by gas stagnation in the small chamber. Section IV details the integration of this apparatus into our experimental setup and presents the results obtained through the implementation of the differential pumping system. Finally, Section V provides a summary of the findings and concludes the paper.\\

\section{Differential pumping with gas jets and high pressure light gases}

\begin{figure}[h!]
	\begin{center}
\includegraphics[width=0.48\textwidth]{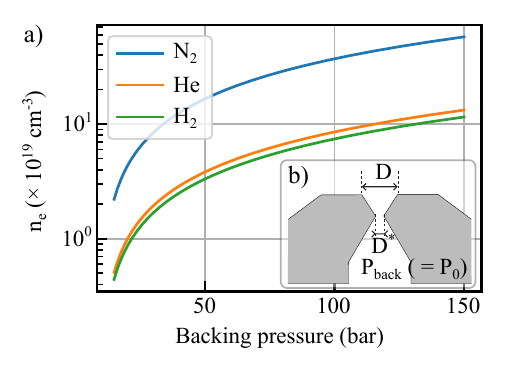}
\caption{a) Electronic density $n_e$ of the plasma $150\,$\textmu m above the gas jet as a function of the backing pressure in the gas jet and the nature of the gas for a 60/180 supersonic nozzle. b) Cross section of the supersonic gas jet, $D^*$ and $D$ are respectively the diameter of the nozzle throat and nozzle opening.}
\label{fig-jet}
\end{center}
\end{figure}

The optimal plasma density and length for a laser-plasma accelerator depend on the physical properties of the laser pulse. In the case of a few mJ, few fs pulse, the plasma density should typically be $n_e\sim 10^{19}-10^{20}\,\mathrm{cm}^{-3}$ and the plasma length $L=100-200\,\text{\textmu}$m \cite{lu_nonlinear_2006, faure_review_2018}. To reach such plasma parameters, we use supersonic cylindrical gas nozzles and backing pressures ranging from a few tens of bar to about $150\,$bar. A cross-section diagram of the gas jet is shown in Figure \ref{fig-jet}b. In the following, we will refer to the gas nozzle parameters as follows: the diameter of the nozzle throat and nozzle opening are respectively written $D^*$ and $D$ and the corresponding throat and opening surfaces are $A^*$ and $A$, where we simply have $A=\pi D^2/4$. For example, a 60/180 gas nozzle has $D^*=60\,\text{\textmu}$m, $D=180\,\text{\textmu}$m, and provides the optimal plasma length and molecular density  $n_{N_2}\sim 10^{19}\,\mathrm{cm}^{-3}$ when using $\mathrm{N_2}$ gas with pressure $P_{back}=30\,$bar backing pressure (see Figure \ref{fig-jet}a). The laser pulse is intense enough to ionize Nitrogen atoms to N$^{5+}$, so that each $\mathrm{N_2}$ molecule releases 10 electrons, leading to the required plasma electron density of $n_e\sim 10^{20}\,\mathrm{cm}^{-3}$.

\subsection{Differential pumping use in high repetition rate LWFA}
\noindent The pumping system is designed so that the residual equilibrium pressure in the experimental chamber, $P_{eq}^{(1)}$ stays sufficiently low and does not affect laser propagation. We set this limit empirically to $P_{eq}^{(1)}<10^{-2}\,$mbar since above this value, turbo-molecular pumps become inoperative, overheat, and the pumping system diverges. 
The nominal pumping speed of the pump, $R_{nom}$, is the theoretical pumping speed advertised by the manufacturer for a perfect pumping system. The effective pumping speed, $R_{eff}$ ($R_{eff} < R_{nom}$), is the real pumping speed in the system. The effective pumping speed is lower than the nominal speed primarily due to the presence of various components between the pump and the jet (belows and valves), which reduce the overall conductance of the pumping system. Additionally, leaks within the system contribute to a reduction in pumping efficiency, and the pumping speed varies with the operating pressure. The residual pressure at equilibrium $P_{eq}^{(1)}$ in the chamber can be estimated knowing the pumping speed $R_{eff}[\mathrm{l/s}]$ and the gas leak rate due to the gas jet $Q[\mathrm{mbar \cdot l/s}]$ by simply writing that:
\begin{equation}\label{eq-leak}
P_{eq}^{(1)}[\mathrm{mbar}]=Q[\mathrm{mbar \cdot l/s}]\ /\ R_{eff}[\mathrm{l/s}]
\end{equation}

\noindent This allows to easily estimate the gas leak rate in our experiment. Typically, in Nitrogen, using a nozzle with $D^*=60\,\text{\textmu}$m and a backing pressure of $P_{back}=20\,$bar, we measure $P_{eq}^{(1)}=2.2\times10^{-3}\,$mbar. Knowing that the pumping speed of our turbo-molecular pump is $R_{eff}=4000\,$l/s for Nitrogen at this pressure, we obtain that the gas leak is $Q=8.8\,\mathrm{mbar\cdot l/s}$.

\noindent Let us now calculate what happens if one uses $\mathrm{H_2}$ instead of Nitrogen. We assume complete ionization: each $\mathrm{H_2}$ molecule now releases two electrons instead of ten in $\mathrm{N_2}$. In addition, we use a 1D isentropic expansion model for compressible gas flows to extrapolate the molecular density in $\mathrm{H_2}$. The molecular density in the flow, $n_{mol}$, can be expressed according to the Boltzmann constant $k_B$, the Mach number $M_{Mach}$ and the initial pressure and temperature, $P_0,\,T_0$ in the reservoir \cite{zucker_fundamentals_2002}: 
\begin{equation}
    n_{mol} = \frac{P_0}{k_BT_0}\left( 1 + \frac{\gamma - 1}{2}M_{Mach}^2\right) ^{-\frac{1}{\gamma - 1}}\\
\end{equation}
\noindent The coefficient $\gamma=c_p/c_v$ is the ratio of specific heat at constant pressure $c_p$ over the specific heat at constant volume $c_v$. For perfect diatomic gases such as $\mathrm{N_2}$ and $\mathrm{H_2}$, $\gamma= 7/5$, whereas for a perfect monoatomic gas, such as $\mathrm{He}$, $\gamma= 5/3$. For our gas jet, where $M_{Mach} \gtrsim 3$, with $P_0 = 20\,$bar and $T_0 = 300\,$K, we obtain the same molecular density at the exit of the nozzle $n_{mol} = 4.3 \times 10^{18}\,\mathrm{cm}^{-3}$ for $\mathrm{N_2}$ and $\mathrm{H_2}$. The resulting electronic density is $n_e = 4.3 \times 10^{19}\,\mathrm{cm}^{-3}$  for $\mathrm{N_2}$ and $n_e = 8.6 \times 10^{18}\,\mathrm{cm}^{-3}$  for $\mathrm{H_2}$. The final electronic density as a function of the backing pressure for different gases is shown in Figure \ref{fig-jet}a. Using $\mathrm{H_2}$, the backing pressure should be increased by a factor of 5 to reach the same electronic density as in $\mathrm{N_2}$, i.e. $P_0=100\,$bar, which means a 5-fold increase in the gas flow. In addition, the leak rate depends on the mass of the molecule as $Q\propto 1/\sqrt{M}$ which is another aggravating factor for the case of $\mathrm{H_2}$ (see next section). Thus, for the same final electron plasma density, the leak rates in $\mathrm{H_2}$ and $\mathrm{N_2}$ are related by the Graham's law of diffusion:
\begin{equation}
Q_{\mathrm{H_2,\,P_0 = 100 bar}}=5Q_{\mathrm{N_2,\,P_0 = 20 bar}}\left( \frac{M_{N_2}}{M_{H_2}} \right)^{1/2}
\end{equation}
\noindent so that instead of $Q=8.8\,\mathrm{mbar\cdot l/s}$ for 20 bar of $\mathrm{N_2}$, the leak rate would be  $Q=165\,\mathrm{mbar\cdot l/s}$ for 100 bar of $\mathrm{H_2}$. To compensate for that increase and keep the background pressure below $P_{eq}^{(1)}=3\times10^{-3}\,$mbar, as when using $\mathrm{N_2}$, the pumping speed should be higher than $R_{eff}=5.5\times10^4\,$l/s which is unrealistic as the largest turbo-molecular pumps on the market have maximum pumping speeds of 4000 l/s. Moreover, one should notice that $\mathrm{H_2}$ is more difficult to pump than $\mathrm{N_2}$. Overall, heavy gases are easier to pump than light gases; the compression ratio of the turbo-molecular pump goes as $K \propto \exp(\sqrt{M})$ with M the molar mass of the gas, and the pumping speed goes as $R \propto -\ln(K)$ \cite{gaede_diffusion_1915, becker_turbomolecular_1966}, so the lighter the gas, the slower the pumping speed of the turbomolecular pump. This has been the bottleneck for using continuous flows of $\mathrm{He}$ or $\mathrm{H_2}$ in LWFA. For example, in our system, the pumping speed of our turbo-molecular pump drops down from $4000\,$l/s in $\mathrm{N_2}$ to $300\,$l/s in $\mathrm{H_2}$ at $P=10^{-1}\,$mbar. In practice, this means that for $P_0=100\,$bar of $\mathrm{H_2}$, the equilibrium pressure would be around $P_{eq}^{(1)}\simeq 0.5\,$mbar, which is likely too high for the turbo-molecular pump to operate. Similar conclusions could be reached for $\mathrm{He}$.

\begin{figure}[htbp]
\begin{center}
\includegraphics[width=0.5\textwidth]{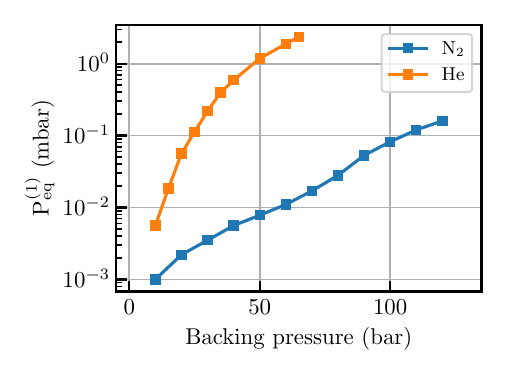}
\caption{Without differential pumping: measured equilibrium pressure in the main chamber versus backing pressure for a gas nozzle with nominal throat diameter $D^*=60\,\text{\textmu}$m (measured throat diameter: $D^*=40\pm2\,\text{\textmu}$m).}
\label{fig-diff1}
\end{center}
\end{figure}

\noindent This has been verified experimentally: we measured the pressure in the vacuum chamber in which a supersonic nozzle lets out a continuous flow of gas. The vacuum chamber is pumped by a $R_{eff}=4000\,$l/s turbo-molecular pump connected to a primary pump with pumping speed $R_{nom} = 130\,$l/s. In Figure \ref{fig-diff1}, we show the variation of the equilibrium pressure versus backing pressure for $\mathrm{N_2}$, and $\mathrm{He}$. The results show very clearly that for $\mathrm{He}$, the equilibrium pressure increases abruptly above $P_{back}=20\,$bar and the pumping system diverges above that, thus precluding operation at high $\mathrm{He}$ flow. This clearly confirms that a specific solution should be implemented for pumping high-pressure ($P>100\,$bar) light gases.

\subsection{Design of the differential pumping system}
\begin{figure}
\begin{center}
\includegraphics[width=0.45\textwidth]{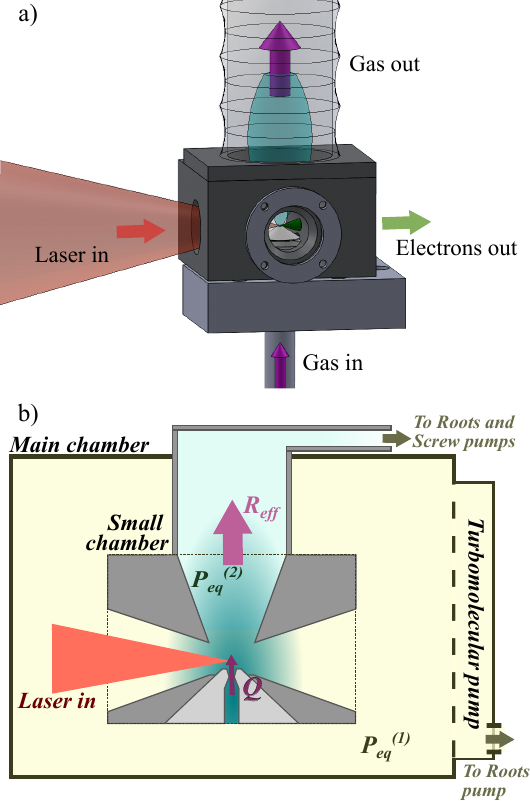}
\caption{a) Schematic of the differential chamber. b) Conceptual design for differential pumping with a free-flowing high-density gas jet.}
\label{fig:art_concept}
\end{center}
\end{figure}

An alternative solution is to use a differential pumping system: the gas jet is enclosed into a small vacuum chamber, the differential chamber, that is pumped by a powerful primary pump able to work efficiently at the mbar level with a pumping speed of $R_{nom} = 360\,$l/s. The idea is to pump out most of the gas from the small chamber while tolerating a higher equilibrium pressure at the mbar level, $P_{eq}^{(2)}\sim\,$mbar within the differential chamber subsystem. Then, by restricting the leakage from the differential chamber to the main chamber, the pressure can be kept below $P_{eq}^{(1)}<10^{-3}\,$mbar in the main chamber. In this case, the laser propagation is not affected by the residual gas, provided that the mbar region in the differential chamber is sufficiently short, for example in our case millimeter scale. Figure \ref{fig:art_concept} shows a conceptual schematic of the differential scheme that is implemented in the experiment. The laser enters and exits the chamber via two cones that terminate with $1\,$mm holes for restricting gas leakage. In addition, the ``high'' density region is limited to a $4$-mm gap in between the two cones to limit its impact on laser propagation.\\

\noindent\emph{1D model used to design the differential pumping}\\
The experimental measurement of the gas leak rate $Q$ can be compared to a theoretical estimation using well-known fluid gas dynamics \cite{zucker_fundamentals_2002}. An estimate of the mass flow rate can be obtained using a 1D isentropic model of the gas flow in the nozzle. For a 1D stationary and turbulent flow, the mass flow rate $\dot{m}= \rho\nu A$ is constant throughout the flow; at the nozzle throat, it reduces to: 
\begin{equation}\label{eq-massflow}
\dot{m}=\rho^*\nu^*A^*=\gamma^{1/2}\left(\frac{\gamma+1}{2}  \right)^{-\frac{\gamma+1}{2(\gamma-1)}}\frac{P_0}{\sqrt{RT_0}}A^*
\end{equation}
where $\rho^*,\,\nu^*$ and $A^*$ are respectively the mass density, fluid velocity, and the surface at the nozzle throat, $P_0$ and $T_0$ are the pressure and temperature in the reservoir. Note that here, the $_0$ subscript refers to the reservoir, i.e. the stagnation volume where the fluid velocity is $\nu_0=0$. The leak rate in $\mathrm{mbar\cdot l/s}$ can then be obtained from $Q\,=PdV/dt=k_BT\frac{dN}{dt}$ which can thus be written as:
\begin{equation}\label{eq-Q}
Q=\dot{m}RT
\end{equation}
with $R=R_0/M$, $R_0=8.314\,$J/mol/K is the ideal gas constant and $M$ is the molar mass of the fluid. Taking $P_{back}=P_0=20\,$bar and $T=T_0=300\,$K for a nozzle with $D^{*}=60\,\text{\textmu}$m, we obtain $Q=11.6\,\mathrm{mbar\cdot l/s}$. The nominal throat diameter of the jets is $D^{*}=60\,\text{\textmu}$m. However, the actual throat diameter deviates from the generic requested value; for the dataset shown in Figure \ref{fig-diff1} the measured throat diameter is $D^{*}=40\pm2\,\text{\textmu}$m, leading to $Q=5.1\,\mathrm{mbar\cdot l/s}$ which is close to the experimental value found earlier ($Q_{exp} = 8.8\,\mathrm{mbar\cdot l/s}$). This analysis validates the fact that this simple model can be used for designing the pumping system as it provides a pressure estimate that matches the experimental measurement within a factor of two.

\begin{figure}
\begin{center}
\includegraphics[width=0.5\textwidth]{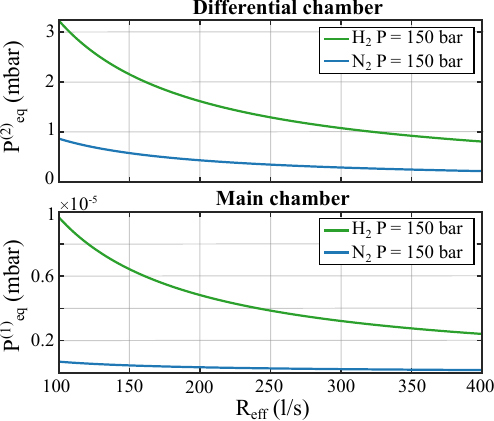}
\caption{Equilibrium pressure in the chambers as a function of effective pumping speed of the differential chamber $R_{eff}$ for a $\mathrm{H_2}$ gas leak of $320\,$mbar.l/s, corresponding to a backing pressure of $150\,$bar and a nozzle with $D^*=60\,\text{\textmu}$m. The pumping speed for the main chamber is assumed to be $4000\,$l/s and the two holes for the differential pumping are $1\,$mm in diameter.}
\label{fig-diff3}
\end{center}
\end{figure}

\begin{figure}[h!]
\begin{center}
\includegraphics[width=0.5\textwidth]{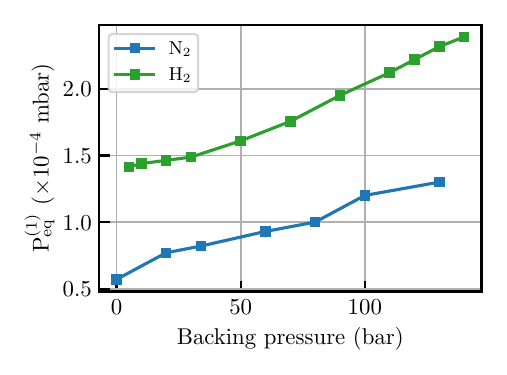}
\caption{With differential pumping: equilibrium pressure in the main chamber versus backing pressure for a gas nozzle with throat diameter $D^*=60\,\text{\textmu}$m. The differential chamber is pumped by a primary pump of  $360\,$l/s.}
\label{fig-diff2}
\end{center}
\end{figure}

\noindent Using this model, it is straightforward to determine $P_{eq}^{(1)}$ and $P_{eq}^{(2)}$, respectively the equilibrium pressures in the experimental and differential chambers, that are reached as a function of the pumping speed $R_{eff}$ used for pumping the differential chamber. Figure \ref{fig-diff3} shows the results for a backing pressure of $P=150\,$bar in a nozzle with throat diameter $D^*=60\,\text{\textmu}$m. It is clear that the effective pumping speed needs to be larger than $R_{eff}>100\,$l/s to maintain the pressure in the differential chamber at the mbar level. When this condition is fulfilled, these estimates confirm that the equilibrium pressure in the main chamber remains at the level $P_{eq}^{(1)}\sim 10^{-5}\,$mbar, where the turbo-molecular pump works in optimal conditions.\\
\noindent Once the differential pumping system was set up, we measured the variation of the equilibrium pressure versus backing pressure for $\mathrm{N_2}$ and $\mathrm{H_2}$. Figure \ref{fig-diff2} shows the pressure measurements in the main chamber with the differential pumping apparatus in operation. The observed discrepancy between the experimental measurements and the theoretical predictions shown in Figure \ref{fig-diff3} arises because the mathematical model inherently simplifies or neglects certain real-world factors that influence the system's performance, such as leaks in the chamber or deviations of the actual conductance from a geometrically simplified model. Experimentally, the lowest pressure achieved in the main chamber without load is approximately $5 \times 10^{-5}\,$mbar, which is significantly higher than the model's prediction of $2.5 \times 10^{-6}\,$mbar. Nevertheless, the equilibrium pressure in the main chamber $P_{eq}^{(1)}$ remains below $3 \times 10^{-4}\,$mbar, even with a backing pressure of $140\,$bar of $\mathrm{H_2}$ in the gas jet.

\section{Numerical simulations}
\subsection{Fluid dynamics simulations}

\begin{figure}
\begin{center}
\includegraphics[height=0.5\textwidth, angle = 0]{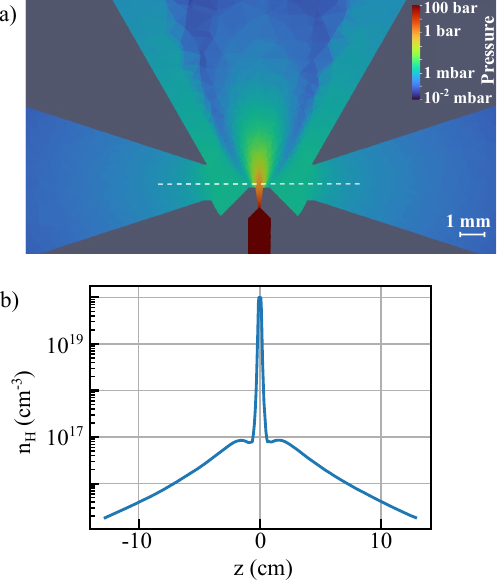}
\caption{a) Results of CFD simulations showing the pressure in the volume of the differential chamber. b) Lineout of the molecular density along the laser axis. }
\label{fig-diff4}
\end{center}
\end{figure}

\noindent In order to obtain more accurate estimates of the gas flow in the nozzle and the differential chamber, we have performed numerical simulations using the computational fluid dynamics software ANSYS Fluent. Simulations were performed using molecular hydrogen modeled by the Aungier-Redlich-Kwong real gas equation. 3D geometry was used. The results are shown in Figure \ref{fig-diff4} for $\mathrm{H_2}$ with a backing pressure $P_{back}=100\,$bar and a nozzle throat of $D^*=60\,\text{\textmu}$m. In the simulations, the boundary conditions at the entrance of the cones and the pumping outlet were taken to be $P=0$. The simulation results shown in Figure \ref{fig-diff4}a confirm that the pressure remains at the mbar level in the differential chamber, away from the region where the high-density jet is present. Figure \ref{fig-diff4}b shows a dramatic decrease of the density by three orders of magnitude in the gap region between the two cones. The density then steadily decreases along the cone which is favorable for coupling the intense laser into the high-density gas jet without prior interaction. Thus, the gas remains well-confined within the differential chamber, but there is still some stagnation observed about 1 cm before the gas jet, potentially impacting the laser pulse quality at the focus.

\subsection{Propagation of the laser pulse up to the gas jet}

\begin{figure}
\begin{center}
\includegraphics[width=0.5\textwidth]{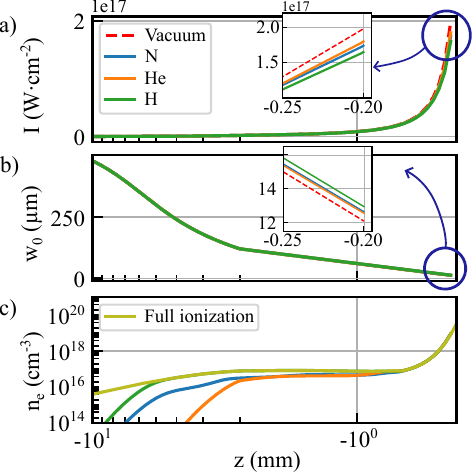}
\caption{Evolution of the laser parameters as it propagates through the gas profile from $z = -10\,$mm to $z = -200\,\text{\textmu}$m. a) Evolution of the laser intensity $I$ in $\mathrm{N}$, $\mathrm{He}$ and $\mathrm{H}$. b) Evolution of the waist of the pulse $w_0$ in the three gases. c)\,Electron density along the propagation in the three gases and the target electron density (corresponding to fully ionized $\mathrm{He}$ or $\mathrm{H}$ gas, and up to the $5^{\mathrm{th}}$ in $\mathrm{N}$ gas). }
\label{fig:par_las}
\end{center}
\end{figure}

\noindent To evaluate whether the stagnating gas affects laser propagation, we performed simulations of the propagation of the laser pulse from the parabola (at  $z = -10\,$cm) to the jet (at $z =  -200\,\text{\textmu}$m) using an optical propagation code. The simulations were conducted using three gases: $\mathrm{H}$, $\mathrm{He}$ and $\mathrm{N}$ in order to assess the potential impact of different gases on the laser pulse. The propagation of the laser through an initially neutral gas was modeled using the open-source library “Axiprop” \cite{andriyash_hightower8083axiprop_2024}. This library uses various optical propagators, depending on the specific conditions (gas or vacuum for instance) and concatenate them to handle more complex scenarios. For our study, the paraxial propagator was used for vacuum propagation from the parabola ($z = -10\,$cm) to the point where the gas density becomes non-negligible ($z = -10\,$mm), as it handles drastic changes in beam size \cite{goodman_introduction_2005}. For the region with gas ($z = -10\,$mm to $z = -200\,\text{\textmu}$m), a non-paraxial propagator was applied, following the method by Oubrerie and al. \cite{oubrerie_axiparabola_2022}. Both propagators were implemented using Hankel transforms to account for cylindrical symmetry \cite{oubrerie_axiparabola_2022}. To model the propagation of the laser in the gas/plasma, the solver that propagates the field within the plasma is similar to the one described by Couairon and al. \cite{couairon_practitioners_2011}. Gas ionization is simulated using the ADK model \cite{ammosov_tunnel_1986, augst_tunneling_1989}. In the gas region, grid sizes were set to $\Delta z = 5.0\,\text{\textmu}$m for $\mathrm{He}$ and $\mathrm{N}$, $\Delta z = 4\,\text{\textmu}$m for $\mathrm{H}$, and $\Delta r = 0.64\,\text{\textmu}$m in all three gases.\\
\noindent The gas density profile was obtained from the Fluent simulations with $\mathrm{N_2}$, renormalized so that the peak electron density reaches $n_e = 2 \times 10^{20}\,\mathrm{cm}^{-3}$ after ionization in all three gases. This adjustment considers that $\mathrm{N}$ contributes 5 electrons, $\mathrm{He}$ contributes 2, and $\mathrm{H}$ contributes 1, allowing us to simulate laser propagation at the same electron density for different gases. The laser parameters were kept constant for each case: a duration of $4.2\,$fs (FWHM), energy of $2.7\,$mJ, and a focal spot size of $4.5\,\text{\textmu}$m (FWHM), matching our experimental parameters.\\
\noindent Figure \ref{fig:par_las} illustrates the evolution of the laser intensity ($I$) and waist ($w_0$) along its propagation through the density profiles of $\mathrm{N_2}$, $\mathrm{He}$, and $\mathrm{H_2}$ gases, with vacuum included for reference. Overall, the laser intensity and beam size remain similar when propagating through vacuum and gas media. However, small differences in laser perturbations become apparent after propagation. Figure \ref{fig:par_las}a shows a reduction in laser intensity when propagating through gas compared to vacuum. This reduction is most pronounced in $\mathrm{H_2}$ (reduction about 19\% at $z = -200\,$\textmu m), and there is a smaller drop of roughly 10\% in $\mathrm{N_2}$ and $\mathrm{He}$. Figure \ref{fig:par_las}b highlights a slight increase in beam waist due to ionization-induced defocusing, with an increase of about 8\% in $\mathrm{H_2}$, and 4\% in $\mathrm{N_2}$ and $\mathrm{He}$ relative to vacuum: along this propagation, $\mathrm{H_2}$ has a greater impact on the laser pulse than He or $\mathrm{N_2}$, though the overall perturbations remain limited.\\

\begin{figure}
\begin{center}
\includegraphics[width=0.45\textwidth]{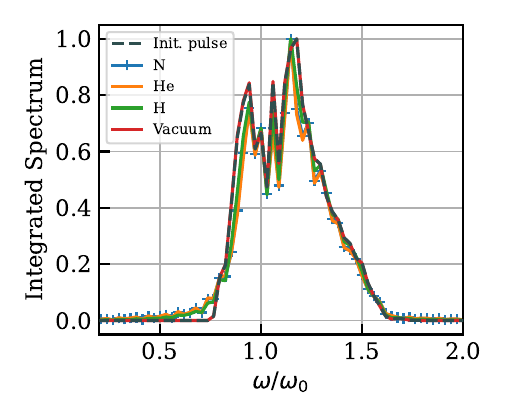}
\caption{Spectral evolution of the laser pulse after propagation in vacuum, and initially neutral $\mathrm{N}$, $\mathrm{He}$ and $\mathrm{H}$ gases.}
\label{fig:blueshift}
\end{center}
\end{figure}

\begin{figure*}[htbp]
    \centering
    \includegraphics[width = \textwidth]{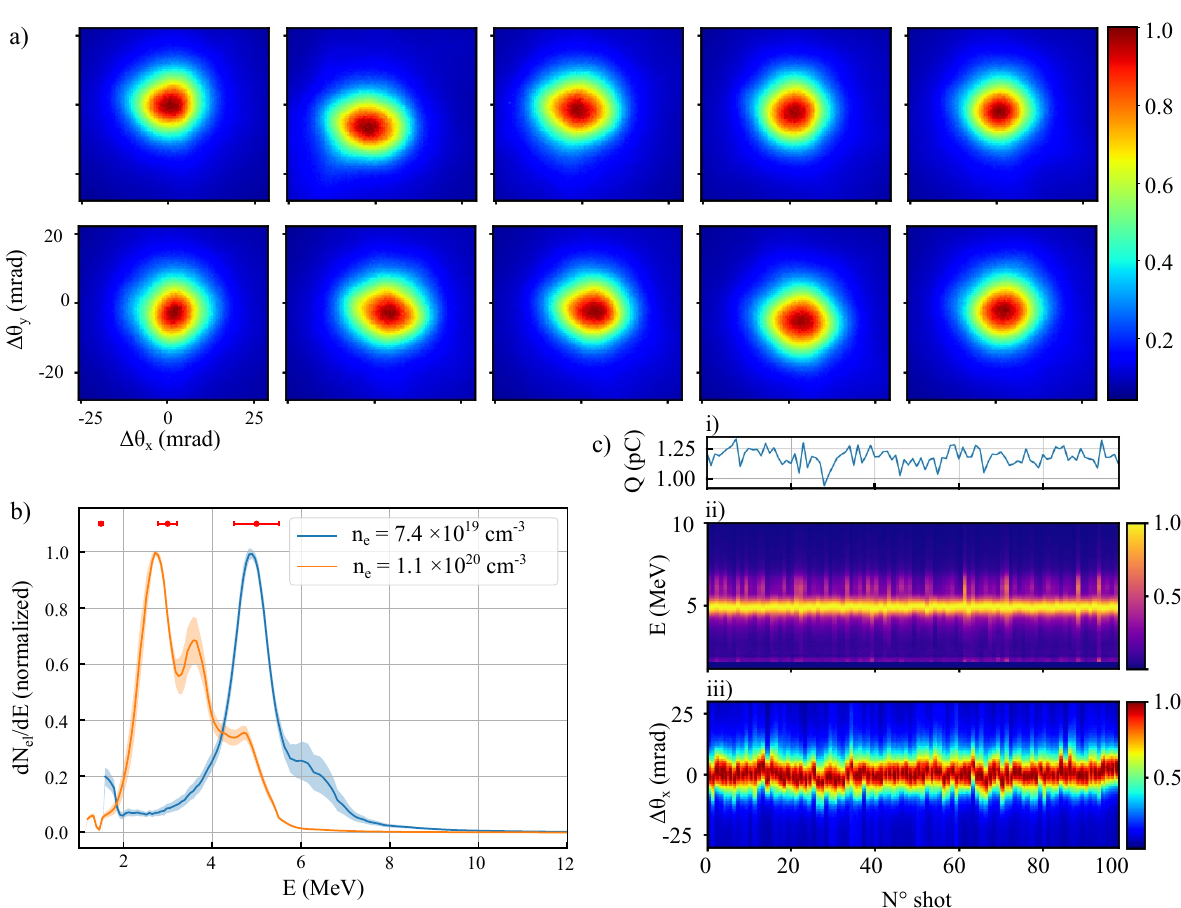}
    \caption{Experimental results obtained in a $\mathrm{H_2}$ (doped $\mathrm{N_2}$) plasma. (a) Single-shot beam profile samples obtained over 10 consecutive shots with $n_e = 1.1 \times 10^{20}\,\mathrm{cm}^{-3}$. (b) Electron spectra. The thickness of the lines corresponds to the RMS fluctuations of the spectrum. (c) Stability over a 100 shots with $n_e = 7.4 \times 10^{19}\,\mathrm{cm}^{-3}$. (i) Charge, (ii) spectra and (iii) beam profile in the $x$ direction, for each shot a slice of 3 pixels around the centroid is shown.}
    \label{fig:res_exp}
\end{figure*}

\noindent While this might initially seem counterintuitive, the key lies in the dynamics of the ionization process. The analysis tracks the evolution of the laser pulse from $z = -10\,$mm to $z = -200\,$\textmu m before the center of the gas jet. Importantly, ionization occurs continuously along the laser propagation path, rather than at a single point. Additionally, the neutral gas density in simulations with $\mathrm{H_2}$ or $\mathrm{He}$ must be five times higher than that of $\mathrm{N_2}$ to maintain consistent electronic density at the focus, since these different molecules release different numbers of electrons upon ionization. As shown in Figure \ref{fig:par_las}c, $\mathrm{H_2}$ molecules are fully ionized from $z = -6\,$mm onward, whereas $\mathrm{N_2}$ molecules are ionized only up to their first level there. As long as the laser ionizes only a single level of $\mathrm{N_2}$, it behaves similarly to when propagating in $\mathrm{H_2}$, but in a lower neutral gas density. On this interval, ionization has a greater impact in $\mathrm{H_2}$ due to its higher neutral gas density. This dynamics shifts around $z = -1\,$mm, where the laser fully ionizes $\mathrm{He}$ and $\mathrm{N_2}$ (up to its fifth level). Up to $z = -1\,$mm, the laser pulse experiences greater degradation in $\mathrm{H_2}$ compared to $\mathrm{He}$ or $\mathrm{N_2}$. Beyond this point, ionization-induced distortions of the laser pulse become more pronounced in $\mathrm{N_2}$, particularly in the main density peak \cite{monzac_optical_2024}, counterbalancing the earlier trends observed during the initial propagation.\\
\noindent Also, the interaction of the pulse with all three gases shifts the spectrum towards the blue. Figure \ref{fig:blueshift} shows the spectrum of the initial pulse and of the pulse at the end of the simulations, integrated over the radial axis, after propagating through vacuum, $\mathrm{N_2}$, $\mathrm{He}$, or $\mathrm{H_2}$. When propagating through the three gases, the spectrum shifts slightly to higher frequencies, with a maximum normalized frequency shift of $\omega_{0,\text{max}}/\omega_0 = 1.02$. The laser pulse behaves similarly in all three gases, with no significant differences regarding the final spectrum.\\
\noindent The stagnating gas has a limited impact on laser propagation. While the interaction of the laser pulse with the gases results in a reduction in intensity, some increase in beam size, and a minor blueshift in the spectrum, these changes do not prevent a priori electron acceleration. The effects are slightly more pronounced in $\mathrm{H_2}$ due to its higher neutral gas density and to the fact that it fully ionizes significantly earlier than $\mathrm{He}$ and $\mathrm{N_2}$.

\section{Implementation of the differential pumping apparatus : experimental results}

After designing and simulating the differential pumping system, we implemented it in our experimental setup to test it in situ, aiming to explore the performance potential of LWFA at 1kHz with this system.

\subsection{Experimental set-up}
\noindent The experimental setup is identical to that described in details in Ref. \cite{monzac_optical_2024}.  The principal information is summarized below. The laser wakefield accelerator is driven by a kHz laser that delivers $4\,$fs pulses at FWHM with $2.7\,$mJ on-target energy \cite{bohle_compression_2014, ouille_relativistic-intensity_2020}. The pulses are focused by a $100\,$mm off-axis parabola down to a $4.5\,\text{\textmu}$m spot at FWHM, reaching a peak intensity in vacuum $I = 1.8 \times 10^{18}\,\mathrm{W \cdot cm}^{-2}$. The laser is focused at a distance of $150\,\text{\textmu}$m from the exit of a continuously flowing supersonic-shocked gas jet with a $180\,\text{\textmu}$m exit diameter \cite{tomkus_high-density_2018, marcinkevicius_femtosecond_2001}. Density characterization is performed with $\mathrm{N_2}$ gas \cite{rovige_symmetric_2021, monzac_optical_2024} using a quadriwave lateral shearing interferometer (SID4-HR, Phasics \cite{primot_achromatic_1995, primot_extended_2000}). The density that we refer to in this paper corresponds to the density $150\,\text{\textmu}$m above the nozzle exit.\\
\noindent Specifically, we used shocked gas jets and Hydrogen doped with 2\% of Nitrogen. Shocked jets helped confine the electron-trapping region, resulting in electron beams with reduced energy spreads and improved stability \cite{bulanov_particle_1998, suk_plasma_2001, tomassini_production_2003, kim_electron_2004}. Additionally, shock-assisted ionization injection demonstrated an increase in trapped charge compared to pure shock injection while maintaining excellent stability  properties \cite{thaury_shock_2004}.\\
\noindent The electron beam charge, spatial profile, and beam pointing were measured using a calibrated YAG screen imaged onto a 14 bits CCD camera. The electron energy spectrum is measured onto the same YAG screen, by inserting a motorized magnetic spectrometer in the beam, which relies on a set of permanent magnetic dipoles creating a field of $0.12\,$T on a $20\,$mm length. Each image presented here was obtained by acquiring data over $1\,$ms (so each image corresponds to a single shot). For each plasma density studied, a series of 100 acquisitions is taken. Statistics over electron beam parameters were then obtained by averaging over the acquisition series, and the uncertainties represent the RMS deviation from the mean value.

\subsection{Results}

\begin{table}
\caption{\label{tab:stab beam}Stability of the beam : RMS fluctuations of the key beam parameters.}
\begin{ruledtabular}
\begin{tabular}{l|c|r}
 $n_e\ (\mathrm{cm}^{-3})$  & $7.4 \times 10^{19}$ & $1.1 \times 10^{20}$\\
\hline
Point. stab. x \hspace{10pt}& 1.46 mrad rms \hspace{5pt}& 1.04 mrad rms\\ 
Point. stab. y \hspace{10pt}& 1.97 mrad rms & 2.26 mrad  rms\\ \rule{0pt}{12pt}
$Q$ & 1.2 pC  & 3.5 pC \\
 & ($\pm 6.0\,\%$ rms)& ($\pm 6.7\,\%$ rms)\\ \rule{0pt}{12pt}
%\hline
$E_{mean}$ & 5.02 MeV  & 3.3 MeV \\
 & ($\pm 2.4\,\%$ rms) & ($\pm 1.8\,\%$ rms)\\ \rule{0pt}{12pt}
%\hline
$E_{spread}$ & 0.99 MeV & 1.44 MeV\\
& ($\pm 26\,\%$ rms) & ($\pm 6.9\,\%$ rms)\\ \rule{0pt}{12pt}
%\hline
$\theta_x$ & 13.0 mrad & 18.2 mrad\\
& ($\pm 10\,\%$ rms) &  ($\pm 13\,\%$ rms)\\ \rule{0pt}{12pt}
%\hline
$\theta_y$ & 12.1 mrad& 16.2 mrad\\
& ($\pm 14\,\%$ rms) & ($\pm 11\,\%$ rms)\\
\end{tabular}
\end{ruledtabular}
\end{table}

For each measurement, we optimized the laser pulse duration and focal position by adjusting the jet position in $5\,$\textmu m steps, while varying the laser pulse chirp in $1.6\,\mathrm{fs^2}$ steps. We explored densities ranging from $n_e = 7.4 \times 10^{19}\,\mathrm{cm}^{-3}$ to $n_e = 1.2 \times 10^{20}\,\mathrm{cm}^{-3}$. Figure \ref{fig:res_exp}a shows a typical electron beam profile at the higher density of $n_e = 1.2 \times 10^{20}\,\mathrm{cm}^{-3}$. The beam exhibits a quasi-circular shape with minimal divergence, as well as a remarkably good shot-to-shot stability, as shown in Table \ref{tab:stab beam}. Figure \ref{fig:res_exp}b shows the corresponding electron spectra for two different densities, averaged over 100 shots. The thickness of the lines corresponds to the RMS fluctuations in the dataset. Both spectra reveal low fluctuations, with well-defined energy peaks, confirming the consistent beam quality. The beam stability is further demonstrated in Figure \ref{fig:res_exp}c, where we present 100 consecutive single-shot acquisitions. The low variance in beam parameters such as charge, divergence, and energy indicates the high level of stability achieved with this setup.
Table \ref{tab:stab beam} summarizes the relative fluctuations of key beam parameters for the two plasma densities. Notably, the beam-pointing fluctuations remain below 2.3 mrad RMS, with charge fluctuations less than 7\%, and energy fluctuations as low as 2\% shot-to-shot, confirming excellent beam stability throughout the experiment.

\section{Conclusion}
\noindent In this paper, we introduced a novel apparatus that combines high-pressure gas jets with a differential pumping system, enabling the continuous operation of a kHz laser-wakefield accelerator utilizing hydrogen plasma generated from a free-flowing gas jet.\\
\noindent The theoretical calculations demonstrate the importance of this system in meeting operational requirements and guide the design process. The chamber pressure was successfully maintained below $3 \times 10^{-4}\,$mbar, even with a backing pressure of up to 140 bar of hydrogen in the 180 µm aperture gas jet, confirming the effectiveness of this vacuum system. Then, fluid simulations demonstrated that the gas behavior within the chamber does not adversely affect laser pulse propagation.\\
\noindent Finally, we conducted laser wakefield acceleration experiments to test the efficacy of the differential pumping setup, yielding highly stable and well-defined electron beams. We hereby demonstrated the first continuous operation at kHz repetition rates with hydrogen, thereby marking a significant advancement in kHz LWFA using light gases.

\begin{acknowledgments}
\noindent This project has received funding from the European Union's Horizon 2020 research and innovation program under grant agreement JRA PRISE no.\,871124 Laserlab-Europe and IFAST under Grant Agreement No 101004730. This project was also funded by the Agence Nationale de la Recherche under Contract No. ANR-20-CE92-0043-01, and benefited from the support of Institut Pierre Lamoure via the “Accélération laser-plasma haute cadence” fund.\\
\noindent This work has benefited from a grant managed by the Agence Nationale de la Recherche (ANR), as part of the program ‘Investissements d’Avenir’ under the reference (ANR-18-EURE-0014).\\
\noindent Financial support from the R\'{e}gion Ile-de-France (under Contract No. SESAME-2012-ATTOLITE) and the Extreme Light Infrastructure-Hungary Non-Profit Ltd (under Contract No. NLO3.6LOA) is gratefully acknowledged. We also acknowledge Laserlab-Europe, Grant No. H2020 EC-GA 654148 and the Lithuanian Research Council under Grant agreement No. S-MIP-21-3.
\end{acknowledgments}

\section*{Data Availability Statement}
The data that support the findings of this study are available from the corresponding author upon reasonable request.

\nocite{*}
\bibliography{pompage_diff}% Produces the bibliography via BibTeX.

\end{document}